\documentstyle[aps,pra,twocolumn,floats]{revtex}

\begin{document}
\draft

\title{Quantum mechanical counterpart of nonlinear optics}

\author{S. Wallentowitz and W. Vogel}

\address{Arbeitsgruppe Quantenoptik, Fachbereich Physik, Universit\"at
  Rostock\\Universit\"atsplatz 3, D-18051 Rostock, Germany}

\date{December 19, 1996}

\maketitle

\begin{abstract}
  Raman-type laser excitation of a trapped atom allows one to realize
  the quantum mechanical counterpart of phenomena of nonlinear optics,
  such as Kerr-type nonlinearities, parametric amplification and
  multi-mode mixing. Additionally, huge nonlinearities emerge from the
  interference of the atomic wave function with the laser waves. They
  lead to a partitioning of the phase space accompanied by a
  significantly different action of the time evolution in neighboring
  phase-space zones. For example, a nonlinearly modified coherent
  ``displacement'' of the motional quantum state may induce strong
  amplitude squeezing and quantum interferences.
\end{abstract}


\narrowtext 

\section{Introduction}
A single atom trapped in a harmonic potential turns out to be a very
well defined object for studying fundamental phenomena of quantum
dynamics. Since the first realization of such a system in an ion trap
by Neuhauser et al.~\cite{first-observation}, the subject has
stimulated much experimental and theoretical work. As has been shown
by Blockley et al.~\cite{jcm}, the laser-assisted coupling between the
internal and external degrees of freedom of a trapped atom can be
described, under appropriate conditions, by a Jaynes-Cummings model.
This allows one to study phenomena we are familiar with from cavity
QED, such as the micromaser dynamics~\cite{Walther-Haroche}, in the
vibronic motion of a trapped atom~\cite{vaser}. Eventually, several
proposals have been published for preparing nonclassical states, such
as squeezed states~\cite{squeezed} and motional number
states~\cite{number}, and successful experiments have been
performed~\cite{njcm-exp,cat-exp}.

The dynamics of a trapped atom, however, not only allows one to
reproduce effects of cavity QED in the quantized motion. When the
spatial extension of the atomic wave-function representing the
center-of-mass motion is no longer small compared with the driving
laser wavelength, nonlinear effects emerge that have no counterpart in
standard nonlinear optics. It has been shown by Vogel and de Matos
Filho that the atom may undergo a vibronic coupling which is very well
described by a nonlinear, multiquantum Jaynes--Cummings
model~\cite{njcm}. Meanwhile this prediction has been confirmed
experimentally~\cite{njcm-exp} and modifications due to micromotion
have been studied~\cite{micromotion}. The nonlinearities in this model
allow to prepare exciting motional quantum states, such as quantum
superpositions of both coherent~\cite{even-odd} and squeezed
states~\cite{Nieto}, nonlinear coherent states~\cite{ncs,ncsg}, pair
coherent states~\cite{Knight1} and pair cat-states~\cite{Knight2}.
Measurement techniques for the full diagnostics of motional quantum
states have been proposed~\cite{measurement} and
realized~\cite{leibfried}.

These outstanding feasibilities render it possible to rise new types
of questions. The nonlinear Jaynes--Cummings model has introduced new
kinds of nonlinearities that substantially modify phenomena we are
familiar with from nonlinear optics, such as multiphoton absorption
and emission. In nonlinear optics, however, other interactions are
known which leave the electronic transitions of the nonlinear medium
almost unchanged. Examples are the Kerr nonlinearity, parametric
interactions and several types of nonlinear wave-mixings. The question
appears as to whether it is possible to realize such phenomena in the
motional dynamics of a single atom, where the trap potential replaces
a cavity used in nonlinear optics.

In the present contribution we propose Raman-type excitations for
inducing various kinds of nonlinear interactions in the quantized
motion of a trapped atom. We consider the quantum mechanical
counterpart of nonlinear optical effects that do not influence the
electronic degrees of freedom of the atomic medium. We show that even
a single degree of freedom of the atomic center-of-mass motion can be
driven in a strongly nonlinear manner. Surprising phenomena are caused
by the interference effects of the atomic wave function with the
driving light waves. They induce a nonlinear partitioning of the phase
space, the action of the time evolution being different in neighboring
phase-space zones.  This partitioning may be used for the generation
of nonclassical effects like amplitude squeezing and quantum
interferences.

The paper is organized as follows. In Sec. II the basic model for the
Raman-induced motional dynamics is introduced and the effective
Hamiltonian for the nonlinear motional interactions is derived.
Section III is devoted to the nonlinear phase-space partitioning
together with the illustration of its effects in simple examples of
motional dynamics. A summary and some conclusions are given in Sec.
IV.

\section{Raman-induced motional dynamics}
Let us consider an atom harmonically bound in a trap. In general the
atom oscillates in the three principal axes of the trap with
frequencies $\nu_i$ ($i=1,2,3$). The trapped atom is driven in a Raman
configuration with two classical laser fields of frequencies
$\omega_L$ and $\omega_L+\Delta$ ($\Delta \!\ll\! \omega_L$), which
are off-resonant with respect to the electronic transitions, see
Fig.~1.
\begin{figure}
  \vspace*{5mm}
  \hspace*{-1.5cm}\input{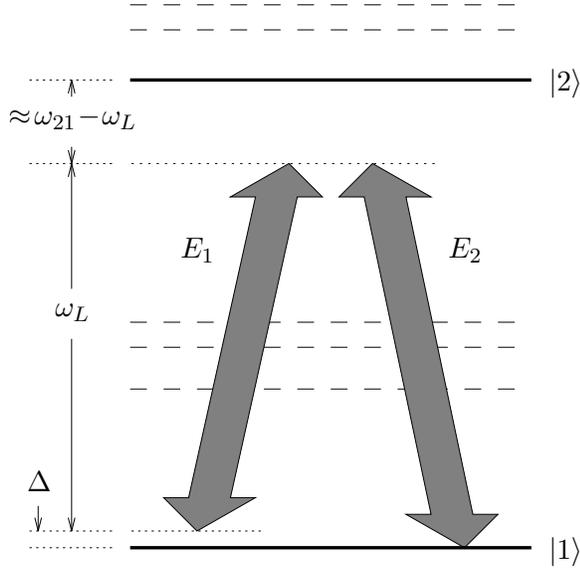}
  \vspace{5mm}
  \caption{The $|1\rangle
    \!\leftrightarrow\! |2\rangle$ transition of a trapped atom is
    driven by two off-resonant laser fields $E_1$ and $E_2$ of
    frequencies $\omega_L$ and $\omega_L\!+\!\Delta$, respectively.
    Other electronic states (broken lines) are far off-resonant. The
    beat frequency $\Delta$ can be tuned on resonance with multiples
    of vibrational frequencies.}
\end{figure}
During the interaction with the two lasers, the atom stays in
its electronic ground state. However, in the resolved sideband regime
and for appropriately chosen laser-beam geometry and laser detuning
$\Delta$, it is possible to affect the motional quantum state of the
atom in a well-controlled manner.

The effective interaction Hamiltonian for the Raman coupling (in
optical rotating-wave approximation) reads as
\begin{equation}
  \label{eq:dipole-hamiltonian}
  \hat{H}_L(t) = \frac{1}{2} \hbar\Omega \, e^{- i \left[\Delta t -
      {\bf k} \cdot \hat{\bf r} \right]} + H.c. ,
\end{equation}
where ${\bf k} \!=\! {\bf k}_1 \!-\! {\bf k}_2$ is the difference
wave-vector of the two laser beams and $\hat{\bf r}$ is the operator
of the atomic center-of-mass position.  For small relative detunings
from the frequency $\omega_{21}$ of the dipole transition
($|\omega_{21}\!-\!  \omega_L|/ \omega_{21}\ll 1$), the effective
two-photon Rabi frequency $\Omega$ is given by
\begin{equation}
  \label{eq:rabis}
  \Omega = \frac{1}{2} \frac{\Omega_1 \, \Omega_2^\ast}{\omega_{21} -
    \omega_L} ,
\end{equation}
with $\Omega_i \!=\! 2 d E_i/\hbar$ ($i \!=\! 1,2$) being the
single-photon Rabi frequencies of the dipole transition of dipole
moment $d$, driven by the electric-field amplitudes $E_1$ and $E_2$ of
the two lasers. The phase of $\Omega \!=\! |\Omega| e^{i\varphi}$ is
determined by the difference phase of the two laser fields $\varphi
\!=\! \varphi_1 \!-\!  \varphi_2$ and can be held very stable in
experiments. Eq.~(\ref{eq:dipole-hamiltonian}) can be written in terms
of creation and annihilation operators of vibrational quanta by using
the relations $k_i \hat{x}_i \!=\! \eta_i ( \hat{a}_i +
\hat{a}_i^\dagger )$, where $k_i$ are the projections of the
wave-vector difference on the principal axes $x_i$ of the trap and
$\eta_i$ are the Lamb--Dicke parameters of the vibration in these
directions. After disentangling the resulting exponential operator
function, the Hamiltonian~(\ref{eq:dipole-hamiltonian}) may be
expanded in a power series as 
\begin{eqnarray}
  \label{eq:expand-hamiltonian}
  \lefteqn{\hat{H}_L(t) = \frac{1}{2} \hbar\Omega \, e^{-i \Delta t}
    \, e^{-(\eta_1^2 + \eta_2^2 + \eta_3^2)/2} \sum_{m m'} \sum_{n n'}
    \sum_{l l'}} & & \\ \nonumber & & \frac{(i\eta_1)^{m+m'}
    (i\eta_2)^{n+n'} (i\eta_3)^{l+l'}}{m! \, m'! \, n! \, n'! \, l! \,
    l'!} \, \hat{a}_1^{\dagger m} \hat{a}_2^{\dagger n}
  \hat{a}_3^{\dagger l} \, \hat{a}_1^{m'} \hat{a}_2^{n'}
  \hat{a}_3^{l'} + H.c.
\end{eqnarray}
This interaction includes, via the mode functions [cf.
Eq.~(\ref{eq:dipole-hamiltonian})] of the laser waves, a
laser-assisted coupling of the three motional degrees of freedom
($x_1, x_2, x_3$). Since the wave-vector difference ${\bf k}$ is
determined by the laser-beam geometry, the coupling of the motional
degrees of freedom can be designed to include one, two, or three
directions.

To consider these couplings in more detail, we assume that the
vibrational frequencies are well resolved by the Raman excitation, so
that we may introduce a vibrational rotating-wave approximation.
Choosing the laser beat frequency to be a multiple of the three
vibrational frequencies, $\Delta \!=\! s_1 \nu_1 \!+\!  s_2 \nu_2$
($s_{1,2}=0,\pm 1, \pm 2, \dots$), one obtains a coupling of all
vibrational modes~\cite{remark1}.  In this case the interaction
Hamiltonian (in the interaction picture) is of the form~\cite{remark2}
\begin{eqnarray}
  \label{eq:ham3d}
  \lefteqn{\hat{H}_{\rm int} =} & & \nonumber \\ \nonumber & &
  \frac{1}{2} \hbar \Omega \sum_{n=-\infty}^\infty \hat{g}_{n\!-\!s_1}
  ( \hat{a}_1^\dagger, \hat{a}_1; \eta_1 ) \, \hat{g}_{n\!-\!s_2} (
  \hat{a}_2^\dagger, \hat{a}_2; \eta_2 ) \, \hat{g}_n (
  \hat{a}_3^\dagger, \hat{a}_3; \eta_3 ) \\ & & + H.c.
\end{eqnarray}
and the operator-valued functions $\hat{g}_k (\hat{a}^\dagger,
\hat{a}; \eta)$ are given by
\begin{equation}
  \label{eq:nlrabi1}
  \hat{g}_k ( \hat{a}^\dagger, \hat{a}; \eta ) = \left\{
    \begin{array}{ll} (i \eta \hat{a}^{\dagger})^{|k|} \, \hat{f}_{|k|} (
      \hat{n}; \eta ) & \quad\mbox{if $k \geq 0$} \\[1ex]
      \hat{f}_{|k|} ( \hat{n}; \eta ) \, (i \eta \hat{a})^{|k|} &
      \quad\mbox{if $k < 0$} \end{array} \right. .
\end{equation}
The Hermitian operator functions $\hat{f}_k(\hat{n}; \eta)$ depend
solely on the number of vibrational quanta $\hat{n}\! =
\!\hat{a}^\dagger \hat{a}$ and read (in normally ordered form) as
\begin{equation}
  \label{eq:nlrabi}
  \hat{f}_k ( \hat{n}; \eta ) = e^{-\eta^2/2} \sum_{l=0}^\infty
  \frac{(-1)^l \eta^{2l}}{l! (l+k)!} \, \hat{a}^{\dagger l} \hat{a}^l
  .
\end{equation}
From Eqs.~(\ref{eq:nlrabi1}) and (\ref{eq:nlrabi}) it is seen, that
for decreasing Lamb--Dicke parameter only the coupling with $k=0$
survives. Therefore, by varying the geometry of the laser-beam
propagation one can vary the Lamb--Dicke parameters in order to change
the Hamiltonian from a coupling of only one, two, or three vibrational
modes.

It is seen from Eqs.~(\ref{eq:ham3d}), (\ref{eq:nlrabi1}) that the
Hamiltonian describes a motional dynamics with the following basic
effects. First, there appear combinations of different powers of the
motional operators $\hat{a}_i$, $\hat{a}_i^\dagger$. Interactions of
this type represent the quantum mechanical counterpart of wave-mixing
effects in nonlinear optics. Second, via the functions
$\hat{f}_k(\hat{n}; \eta)$ the couplings depend in a nonlinear manner
on the excitations of the modes. This results from the interference of
the atomic (center-of-mass) wave functions and the beat node of the
laser waves, which is a typical effect of quantized atomic motion.

\section{Nonlinear phase-space partitioning}
To get some insight in these effects, we first consider the
one-dimensional dynamics, where only the motion in $x_1$-direction is
affected by the lasers ($\eta_2\!=\!\eta_3\!=\!0$). This requires a
geometry of laser propagations with vanishing projections of the
difference wave-vector ${\bf k}$ on the axes $x_2$ and $x_3$. In this
case the Hamiltonian simplifies as
\begin{equation}
  \label{eq:ham1d}
  \hat{H}_{\rm int} = \frac{1}{2} \hbar \Omega \, \hat{f}_k
  \left(\hat{n}; \eta \right) \, (i \eta \hat{a})^k + H.c. ,
\end{equation}
where we assumed a laser detuning of $\Delta \!=\! k \nu_1$ ($k
\!\geq\! 0$) and we have omitted the indices of the $x_1$ direction.
Interactions of this type may be considered as nonlinear mode
couplings of one (weakly excited) quantized mode with (strongly
excited) classical modes. Such approximations are frequently used in
quantum optics. Experiments of the type proposed here would allow to
realize these couplings almost perfectly and to study the additional
(excitation-dependent) nonlinearities.

For example, let us consider the one-quantum resonance
($\Delta\!=\!\nu_1$) in more detail. In this case the structure of the
unitary time-evolution operator obtained from the
Hamiltonian~(\ref{eq:ham1d}) shows some formal resemblance to a
nonlinearly modified coherent ``displacement''
operator~\cite{displace},
\begin{eqnarray}
  \label{eq:evoldisp}
  \hat{U}_{\rm int}(t) & = & \hat{D}\left[ -\frac{\eta \Omega^\ast
      t}{2} \hat{f}_1(\hat{n}; \eta) \right] \\ \nonumber & = &
  \exp\left[ - \frac{\eta \Omega^\ast t}{2} \hat{a}^\dagger \,
    \hat{f}_1(\hat{n}; \eta) + \frac{\eta \Omega t}{2}
    \hat{f}_1^\dagger(\hat{n}; \eta) \, \hat{a} \right] .
\end{eqnarray}
For small values of the Lamb--Dicke parameter, $\eta \!\ll\! 1$,
according to Eq.~(\ref{eq:nlrabi}) the operator~(\ref{eq:evoldisp})
may be replaced by the usual displacement operator $\hat{D}(-\eta
\Omega^\ast t / 2)$.

The nonlinear dependence of the ``displacement''
operator~(\ref{eq:evoldisp}) on the mean number of vibrational quanta
leads to effects of a new type. For a first insight we may replace the
number operator by its eigenvalue. We arrive at the $c$-number
function $f_1(n; \eta) \!=\! \langle n | \hat{f}_1(\hat{n}; \eta) | n
\rangle$, which reads as
\begin{equation}
  f_1(n; \eta) = \frac{e^{-\eta^2/2}}{n+1} L_n^{(1)}(\eta^2) ,
\end{equation}
with $L_n^{(k)}(x)$ being Laguerre polynomials. To consider the action
of the nonlinear displacement in phase space, it is advantageous to
introduce the (complex) phase-space amplitude $\alpha$ by setting $n
\!=\! |\alpha|^2$. The resulting function $f_1(|\alpha|^2; \eta)$ has
zeros and changes its sign for certain values of $|\alpha|$.
Consequently, the direction of the displacement can be reversed,
depending on the amplitude of the quantum state in phase space. That
is, the phase space is effectively partitioned in zones. The action of
the displacement in adjacent zones differs in the fact that the
directions of displacements are opposite to each other, along an axis
which is controlled by the phase difference of the lasers. These
phase-space zones are separated by the circles on which the coupling
function $f_1(|\alpha|^2; \eta)$ changes its sign. This nonlinear
partitioning of the phase space leads to striking consequences with
respect to the evolution of the quantum state.

Let us consider the evolution of a coherent state that is initially
located on the boundary between two such phase-space zones. Inside the
corresponding circle the coupling $f_1(|\alpha|^2; \eta)$ is positive
and outside it is negative. Due to this fact the nonlinear
``displacement'' operator tends to split the coherent state as shown
in Fig.~2.
\begin{figure}
  \input{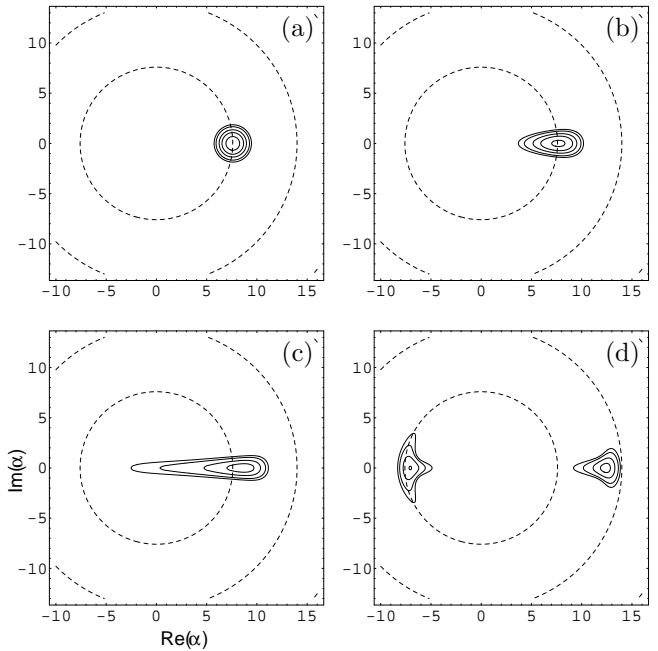}
  \caption{Time evolution of a coherent state that is initially
    placed on the boundary between two phase-space zones with opposite
    displacement directions (chosen along the real axis). The
    dimensionless times $\eta |\Omega| t$ are given by $0$ (a), $2.5$
    (b), $5$ (c), and $15$ (d); $\eta \!=\! 0.25$. The contours
    represent the $Q$~functions of the motional quantum states.}
\end{figure}
For rather short times the state can exhibit a significant reduction
of phase fluctuations. In the further course of time the states is
splitted into well separated substates. This leads to a coherent
superposition of two quantum states, accompanied by
quantum-interference effects. The displacement of each substate is
limited by the boundaries between the phase-space zones, where the
strength of displacement becomes negligible. The result is a squeezing
of each substate onto the corresponding circle partitioning the phase
space.

This effect can be used to generate quantum states exhibiting strong
amplitude squeezing. Let us consider the nonlinear displacement of a
coherent state that is initially located within a single phase-space
zone. As expected, the state is displaced in a well defined direction
in phase space until it is squeezed onto the next circle separating
two zones. The result consists in a strongly amplitude-squeezed
state~\cite{remark5} with a non-vanishing coherent amplitude as shown
in Fig.~3.
\begin{figure}
  \vspace*{-18mm}
  \hspace*{-2mm}\input{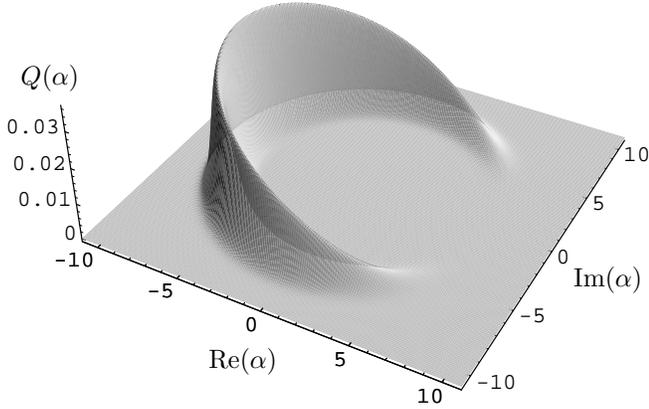}
  \vspace{-10mm}
  \caption{$Q$~function of a strongly amplitude-squeezed state with
    $\langle \Delta \hat{n}^2 \rangle / \langle \hat{n} \rangle \!=\!
    0.006$. This state is reached from an initially coherent state
    ($\alpha \!=\! -9$) in a dimensionless time $\eta |\Omega| t
    \!\approx\! 10$, for $\eta \!=\! 0.25$. The displacement acts
    along the real axis.}
\end{figure}
It is worth noting that in its further evolution this quantum state
does not approach a Fock state. The reason consists in the fact that
in general the transitions between neighboring phase-space zones are
very weak, but not suppressed completely. This leads to continued
deformations of the phase-space distributions of the motional quantum
state.

The one-dimensional Hamiltonian~(\ref{eq:ham1d}) allows to consider
other types of phenomena known from nonlinear optics. Choosing
$k\!=\!0$, the corresponding dynamics is related to the Kerr
nonlinearity~\cite{remark4}. The standard Kerr nonlinearity is
reproduced by expanding the Hamiltonian up to $\eta^4$.  In the more
general case of larger Lamb--Dicke parameters the nonlinear function
$f_0(n; \eta) \!=\! \langle n | \hat{f}_0(\hat{n}; \eta) | n \rangle$
plays a similar role as the function $f_1(n; \eta)$ for the case
$k\!=\!1$. Its oscillations as a function of $n$ again lead to the
phase-space partitioning effect. This is illustrated in Fig.~4 for an
initially coherent state situated at a circle in phase space where
$f_0(n; \eta) = 0$.
\begin{figure}
  \input{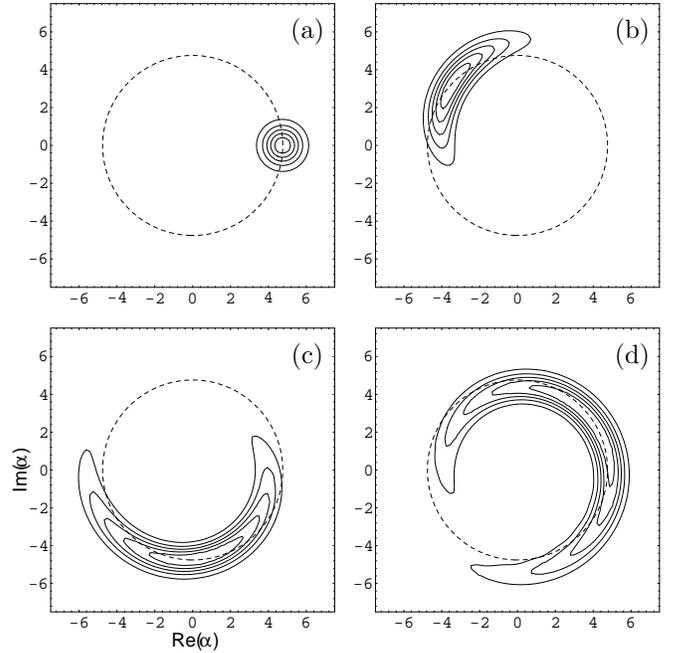}
  \caption{Time evolution of the Q~function for $k\!=\!0$ (Kerr-type
    effects) and $\eta\!=\!0.25$. The dimensionless times $|\Omega| t$
    are: $0$ (a), $173.5$ (b), $346.6$ (c), $500$ (d).}
\end{figure}
One clearly observes a rotation of the state which is due to the term
$\propto\eta^2$ of $\hat{f}_0(\hat{n}; \eta)$.  Moreover, the state is
significantly deformed: inside and outside the circle the state
undergoes phase shifts into opposite directions, reflecting the change
in sign of the coupling.

For $k\!=\!2$ the Hamiltonian~(\ref{eq:ham1d}) represents the
nonlinear generalization of a classically driven parametric
interaction. For $\eta \!\ll\! 1$ the time-evolution operator agrees
with the squeeze operator. This limiting case has been realized
experimentally~\cite{njcm-exp}. In the more general case of larger
Lamb--Dicke parameters, a rather complex dynamics appears. The
interpretation of all of its features needs some further research.

For studying a quantized version of the parametric interaction, the
coupling of two degrees of freedom is needed. Consider a laser-beam
geometry with the projection of the difference wave-vector ${\bf k}$
on the $x_3$-axis being zero, so that $\eta_3\!=\!0$. The dynamics
couples the motion in $x_1$ and $x_2$ directions. For example, a
detuning of $\Delta \!=\! 2 \nu_1 \!-\! \nu_2$ ($s_1\!=\!2,
s_2\!=\!-1$) reduces the interaction Hamiltonian~(\ref{eq:ham3d}) to
\begin{equation}
  \label{eq:ham2d}
  \hat{H}_{\rm int} = -\frac{i}{2} \hbar \eta_1^2 \eta_2 \Omega \,
  \hat{f}_2( \hat{n}_1; \eta_1 ) \, \hat{a}_1^2 \hat{a}_2^\dagger \,
  \hat{f}_1( \hat{n}_2; \eta_2 ) + H.c.,
\end{equation}
representing a nonlinear generalization of the parametric interaction.
For small Lamb--Dicke parameters, $\eta_{1,2} \!\ll\! 1$, this
interaction simplifies to
\begin{equation}
  \label{eq:ham2pc}
  \hat{H}_{\rm int} = -\frac{i}{2} \hbar \eta_1^2 \eta_2 \Omega \,
  \hat{a}_1^2 \hat{a}_2^\dagger + H.c.,
\end{equation}
which is the standard form of the parametric coupling. Beyond the
Lamb--Dicke regime the interaction includes nonlinearities of the type
considered above, which now appear in both motional degrees of
freedom.  Consequently, the nonlinear phase-space partitioning effects
considered above will be of relevance for each degree of freedom
involved in the Raman-induced motional dynamics.

\section{Summary and Conclusions}
In conclusion we have shown, that a Raman-type laser excitation allows
to induce nonlinear interactions of motional degrees of freedom of a
trapped atom, which are closely related to phenomena of nonlinear
optics that do not change the electronic quantum states of the medium.
The number of coupled modes can be easily controlled by the laser beam
geometry. Standard effects can be realized, including coherent
displacements, Kerr nonlinearities, and parametric mode couplings.  In
the laser-assisted motional dynamics additional nonlinearities emerge,
which are caused by the interference between the light waves and the
wave function representing the atomic center-of-mass motion.

An important consequence of these nonlinearities consists in a
partitioning of the motional phase space, which is caused by an
oscillatory behavior of the motional interactions as a function of the
phase-space amplitude. In neighboring phase-space zones the actions of
the time evolution appear to be significantly different from each
other. For example, in two adjacent zones a nonlinearly modified
''displacement'' operator acts in opposite directions. Consequently, a
quantum state whose initial location is on the boundary between two
zones will be splitted in two substates, which eventually gives rise
to quantum interferences. Moreover, the partitioning allows to
generate strongly amplitude-squeezed motional states. Eventually, in
the case of a generalized Kerr nonlinearity the phase-space
partitioning may lead to pronounced deformations of the initial state,
which are caused by opposite phase shifts appearing in adjacent
phase-space zones.

The phase-space partitioning, although illustrated in this paper for
the motional dynamics in one dimension, is a universal feature of the
interference between the Raman beat node and the wave function
describing the center-of-mass motion of the atom. When two or three
dimensions are involved in the Raman-induced dynamics, the
partitioning effects appear in the phase space of each motional degree
of freedom.  Consequently, the coupling between different motional
modes will be strongly influenced by the interplay of these nonlinear
effects. In general the dynamics will sensitively depend on the
initial conditions. Besides the feasibility of realizing phenomena
well known from nonlinear optics in the motion of a trapped atom, this
opens novel possibilities for studying nonlinear phenomena in a
well-defined quantum system.

\section*{Acknowledgments}
This research was supported by the Deutsche Forschungsgemeinschaft.
The authors gratefully acknowledge valuable comments by P.L. Knight
and R.L. de Matos Filho.


\begin{references}
\bibitem{remark1} Since $\nu_3 \!\approx\! \nu_1 \!+\! \nu_2$ holds
  for a quadrupole trapping potential, this choice describes all
  possible combinations of the three vibrational frequencies.
\bibitem{remark2} We consider here the nondegenerate case $\nu_1
  \!\neq\! \nu_2$. For the degenerate case only one vibrational
  frequency is needed to completely describe a resonant detuning:
  $\Delta \!=\! s \nu_1$. In this case the Hamiltonian is given by
  Eq.~(\ref{eq:ham3d}) with an additional sum ($\sum_{s_1 = -
    \infty}^\infty$) over $s_1$ while replacing $s_2 \rightarrow s -
  s_1$. This sum reflects the manifold of possible resonances,
  generated by the degeneracy $\nu_1 \!=\! \nu_2$.
\bibitem{displace} We would like to point out that the nonlinear
  ``displacement'' operator defined in Eq.~(\ref{eq:evoldisp}) does
  not exhibit the standard properties of a displacement operator,
  which is a consequence of the fact that the operator-valued function
  $\hat{f}_1$ does not commute with the annihilation and creation
  operators. However, its action shows some resemblance to a nonlinear
  displacement which depends on the amplitude of the quantum state in
  phase space.
\bibitem{remark5} Note that a hint on such an amplitude-squeezing
  effect has already been found experimentally, cf. footnote~36 of
  Ref.~\cite{cat-exp}.
\end{references}
\end{document}